\documentclass{jnmp}

\usepackage{amsmath}

\begin{document}

\renewcommand{\evenhead}{S Yu Sakovich}
\renewcommand{\oddhead}{On Integrability of Differential Constraints}

\thispagestyle{empty}

\FirstPageHead{9}{1}{2002}{\pageref{firstpage}--\pageref{lastpage}}{Letter}

\copyrightnote{2002}{S Yu Sakovich}

\Name{On Integrability of Differential Constraints Arising from
the Singularity Analysis}

\label{firstpage}

\Author{S Yu SAKOVICH}

\Address{Institute of Physics, National Academy of Sciences, 220072 Minsk, Belarus\\
~~E-mail: saks@pisem.net}

\Date{Received June 18, 2001; Revised June 26, 2001;
Accepted July 30, 2001}

\begin{abstract}
\noindent
Integrability of differential constraints arising from
the singularity analysis of two (1+1)-dimensional second-order
evolution equations is studied. Two nonlinear ordinary
differential equations are obtained in this way, which are
integrable by quadratures in spite of very complicated branching
of their solutions.
\end{abstract}

\noindent
The Weiss-Kruskal algorithm of singularity analysis \cite{WTC},
\cite{JKM} is an effective tool for testing the integrability of
PDEs. Certainly, most of PDEs do not pass the Painlev\'{e} test.
If the test fails at its first step, when the leading exponents
are determined, it is often possible to improve the dominant
behavior of solutions by a transformation of dependent variables
and then continue running the test. If the test fails at its
second step, when the positions of resonances are determined, one
still may hope to improve those positions by a transformation of
hodograph type \cite{S} and continue the analysis again. But if
the test fails at its third step, when the consistency of
recursion relations is checked, one only have to introduce some
logarithmic terms into the expansions of solutions, and this is
generally believed to be a clear symptom of nonintegrability of
the tested PDE \cite{AC}.

If the recursion relations are not consistent at a resonance,
there arise some differential constraints on those arbitrary
functions which have entered the singular expansions at lower
resonances. A conjecture exists, formulated by Weiss \cite{W},
that the differential constraints, arising in the singularity
analysis of nonintegrable equations, are always integrable
themselves. In the present paper, we study this interesting
conjecture on the basis of the
following two (1+1)-dimensional second-order evolution equations:%
\begin{equation}
u_{t}=u_{xx}+uu_{x}-u^{3},\label{e1}%
\end{equation}%
\begin{equation}
u_{t}=u_{xx}-u^{2}.\label{e2}%
\end{equation}

Let us study the equation (\ref{e1}) first. Starting the
Weiss-Kruskal algorithm, we determine that a hypersurface
$\phi(x,t)=0$ is non-characteristic for (\ref{e1}) if
$\phi_{x}\neq0$ (we set $\phi_{x}=1$), and that the general
solution of (\ref{e1}) depends on two arbitrary functions
of one variable. Then we substitute the expansion%
\begin{equation}
u=u_{0}(t)\phi^{\alpha}+\cdots+u_{n}(t)\phi^{n+\alpha}+\cdots\label{expu}%
\end{equation}
into (\ref{e1}) and find admissible branches and positions of
resonances in them. Besides the Taylor expansions with
$\alpha=0,1$, governed by the Cauchy-Kovalevskaya theorem, the
equation (\ref{e1}) admits the following two
singular branches:%
\begin{equation}
\alpha=-1,\quad u_{0}=1,\quad(n+1)(n-3)=0;\label{e1u1}%
\end{equation}%
\begin{equation}
\alpha=-1,\quad u_{0}=-2,\quad(n+1)(n-6)=0;\label{e1u2}%
\end{equation}
both being generic.

In the case (\ref{e1u1}), the coefficients of (\ref{expu}) are
determined by
the recursion relations%
\begin{gather}
(n-1)(n-2)u_{n}+\frac{1}{2}(n-2)\sum_{i=0}^{n}u_{i}u_{n-i}-\sum_{i=0}^{n}%
\sum_{j=0}^{n-i}u_{i}u_{j}u_{n-i-j}-\nonumber\\
(n-2)\phi^{\prime}u_{n-1}-u_{n-2}^{\prime}=0,\quad n=0,1,2,\ldots,\label{rr}%
\end{gather}
where $u_{-2}=u_{-1}=0$ formally, and the prime denotes the
derivative with respect to $t$. We find from (\ref{rr}) that
$u_{1}=\frac{1}{4}\phi^{\prime}$ at $n=1$,
$u_{2}=-\frac{1}{16}\phi^{\prime2}$ at $n=2$, whereas $u_{3}(t)$
is not determined and the differential constraint (or
``compatibility
condition'') on $\phi$,%
\begin{equation}
\phi^{\prime\prime}=\frac{1}{2}\phi^{\prime3},\label{dc1}%
\end{equation}
arises at $n=3$. Consequently, the equation (\ref{e1}) does not
pass the Painlev\'{e} test for integrability. We see, however,
that the differential constraint (\ref{dc1}) is integrable by
quadratures, and this supports the conjecture of Weiss.

For the branch (\ref{e1u2}), we have the same recursion relations
(\ref{rr}) and obtain $u_{1},\ldots,u_{5}$ from them. Then, at
$n=6$, $u_{6}(t)$ remains
undetermined, but the following differential constraint on $\phi$ arises:%
\begin{equation}
\phi^{\prime}\phi^{\prime\prime\prime}+\frac{4}{9}\phi^{\prime\prime
2}-\frac{173}{45}\phi^{\prime3}\phi^{\prime\prime}+\frac{142}{225}\phi
^{\prime6}=0.\label{dc2}%
\end{equation}
By introducing the new variable $v(t)=\phi^{\prime}$, we rewrite
(\ref{dc2})
as the second-order ODE%
\begin{equation}
vv^{\prime\prime}+\frac{4}{9}v^{\prime2}-\frac{173}{45}v^{3}v^{\prime
}+\frac{142}{225}v^{6}=0.\label{v1}%
\end{equation}
Is this ODE integrable? Let us try to answer this question, using
the Ablowitz-Ramani-Segur algorithm of the singularity analysis
of ODEs \cite{ARS}, the predecessor of the Weiss-Kruskal
algorithm. Substituting into
(\ref{v1}) the expansion%
\begin{equation}
v=v_{0}\psi^{\beta}+\cdots+v_{n}\psi^{n+\beta}+\cdots\label{expv}%
\end{equation}
with $\psi^{\prime}=1$ and constant coefficients $v_{i}$,
$i=0,\ldots ,n,\ldots$, we find the following singular branches
and positions of
resonances in them:%
\begin{equation}
\beta=\frac{9}{13},\quad\forall v_{0}:v_{0}\neq0,\quad(n+1)n=0;\label{v1b1}%
\end{equation}%
\begin{equation}
\beta=-\frac{1}{2},\quad v_{0}^{2}=-\frac{5}{2},\quad\left(  n+\frac{37}%
{6}\right)  (n+1)=0;\label{v1b2}%
\end{equation}%
\begin{equation}
\beta=-\frac{1}{2},\quad
v_{0}^{2}=-\frac{155}{284},\quad(n+1)\left(
n-\frac{1147}{852}\right)  =0.\label{v1b3}%
\end{equation}
We see that the ODE (\ref{v1}) does not possess the Painlev\'{e}
property. In principle, the rational values of $\beta$ and $n$
(\ref{v1b1})--(\ref{v1b3}) allow the equation (\ref{v1}) to
possess the weak Painlev\'{e} property, but we will not study
this possibility because the weak Painlev\'{e} property not
always implies integrability \cite{AC}.

Before proceeding with the integrability of (\ref{v1}), let us
obtain one more ODE, with even worse analytic properties, from
the PDE (\ref{e2}). Using the expansion (\ref{expu}), we find
that the equation (\ref{e2}) admits the
singular generic branch%
\begin{equation}
\alpha=-2,\quad u_{0}=6,\quad(n+1)(n-6)=0.\label{e2u}%
\end{equation}
Then, constructing the recursion relations and checking their
consistency at
the resonance $n=6$, we obtain the differential constraint%
\begin{equation}
\phi^{\prime}\phi^{\prime\prime\prime}+\frac{3}{8}\phi^{\prime\prime
2}-\frac{27}{20}\phi^{\prime3}\phi^{\prime\prime}+\frac{9}{200}\phi^{\prime
6}=0.\label{dc3}%
\end{equation}
Consequently, the PDE (\ref{e2}) does not pass the Painlev\'{e}
test for integrability. Again, we rewrite (\ref{dc3}) by
$v(t)=\phi^{\prime}$ in the
form%
\begin{equation}
vv^{\prime\prime}+\frac{3}{8}v^{\prime2}-\frac{27}{20}v^{3}v^{\prime}%
+\frac{9}{200}v^{6}=0,\label{v2}%
\end{equation}
use the expansion (\ref{expv}), and find for the ODE (\ref{v2})
the following
singular branches and positions of resonances in them:%
\begin{equation}
\beta=\frac{8}{11},\quad\forall v_{0}:v_{0}\neq0,\quad(n+1)n=0;\label{v2b1}%
\end{equation}%
\begin{equation}
\beta=-\frac{1}{2},\quad v_{0}^{2}=\frac{5}{2}\left(
-3-\sqrt{6}\right) ,\quad\left(  n+\frac{27}{8}\left(
2+\sqrt{6}\right)  \right)
(n+1)=0;\label{v2b2}%
\end{equation}%
\begin{equation}
\beta=-\frac{1}{2},\quad v_{0}^{2}=\frac{5}{2}\left(
-3+\sqrt{6}\right) ,\quad (n+1)\left(  n+\frac{27}{8}\left(
2-\sqrt{6}\right)  \right)
=0.\label{v2b3}%
\end{equation}
Like the equation (\ref{v1}), the ODE (\ref{v2}) does not possess
the Painlev\'{e} property. But, unlike the ODE (\ref{v1}), which
can possess the weak Painlev\'{e} property, the equation
(\ref{v2}) is characterized by irrational positions of
resonances, i.e.\ its solutions exhibit some infinite branching.

Thus, due to the results of the Painlev\'{e} test, we should not
expect the equations (\ref{v1}) and (\ref{v2}) to be integrable.
Nevertheless, these two ODEs turn out to be integrable by
quadratures.

Let us consider the ODE%
\begin{equation}
vv^{\prime\prime}+av^{\prime2}+bv^{3}v^{\prime}+cv^{6}=0\label{ode}%
\end{equation}
with constant $a,b,c$. After the change of variables%
\begin{equation}
\left\{  t,v(t)\right\}  \rightarrow\left\{  v,w(v)\right\}
:v^{\prime
}=w(v),\label{vw}%
\end{equation}
which can be inverted as $t=\int\frac{\mathrm{d}v}{w(v)}$, the
ODE (\ref{ode})
becomes%
\begin{equation}
vww_{v}+aw^{2}+bv^{3}w+cv^{6}=0.\label{tf1}%
\end{equation}
Changing the variables again,%
\begin{equation}
\left\{  v,w(v)\right\}  \rightarrow\left\{  y,z(y)\right\}
:v=f(y),\,w=g(y)z(y),\label{yz}%
\end{equation}
we rewrite (\ref{tf1}) in the form%
\begin{equation}
zz_{y}+\left(  \frac{g_{y}}{g}+\frac{af_{y}}{f}\right)  z^{2}+\frac{bf^{2}%
f_{y}}{g}z+\frac{cf^{5}f_{y}}{g^{2}}=0.\label{tf2}%
\end{equation}
Then, if $a\neq-3$, we choose%
\begin{equation}
f=\exp\left(  -\frac{y}{a+3}\right)  ,\quad g=\epsilon f^{3},\quad
\epsilon=\mathrm{constant}\neq0\label{fg}%
\end{equation}
in (\ref{tf2}) and obtain the ODE%
\begin{equation}
zz_{y}-z^{2}-\frac{b}{\epsilon(a+3)}z-\frac{c}{\epsilon^{2}(a+3)}%
=0,\label{tf3}%
\end{equation}
which is integrable by quadrature. In the case of (\ref{v1}), setting%
\begin{equation}
f=\exp\left(  -\frac{9}{31}y\right)  ,\quad
g=-\frac{1}{155}\exp\left(
-\frac{27}{31}y\right)  ,\label{fg1}%
\end{equation}
we have%
\begin{equation}
y=\int\frac{z\,\mathrm{d}z}{(z+31)(z+173)}.\label{in1}%
\end{equation}
In the case of (\ref{v2}), setting%
\begin{equation}
f=\exp\left(  -\frac{8}{27}y\right)  ,\quad
g=\frac{1}{15}\exp\left(
-\frac{24}{27}y\right)  ,\label{fg2}%
\end{equation}
we have%
\begin{equation}
y=\int\frac{z\,\mathrm{d}z}{z^{2}-6z+3}.\label{in2}%
\end{equation}
The integrals (\ref{in1}) and (\ref{in2}) reveal the origin of
the observed complicated branching of solutions of (\ref{v1}) and
(\ref{v2}).

Consequently, the differential constraints (\ref{dc1}),
(\ref{dc2}) and (\ref{dc3}) are integrable by quadratures, and
this supports the conjecture of Weiss. This interesting
conjecture deserves further study based on more examples of
nonintegrable PDEs.

The ODEs (\ref{v1}) and (\ref{v2}) may illustrate the fact that
the analytic properties of integrable equations can be very
complicated. In this relation, the following two papers should
also be mentioned. Lemmer and Leach \cite{LL} studied the class
of ODEs $Y^{\prime\prime}+YY^{\prime}+KY^{3}=0$ with constant $K$
(this class is equivalent to (\ref{ode}) with $a=1$) and found
that the Lie integrability in quadratures is wider than the
Painlev\'{e} integrability. More recently, Ramani, Grammaticos
and Tremblay \cite{RGT} found that the Painlev\'{e} property is
not necessary for integrability of a large class of linearizable
systems.

\label{lastpage}

\end{document}